\documentclass[conference]{IEEEtran}
\usepackage{pgfplots, pgfplotstable}
\usepackage{makecell}
\pgfplotsset{compat=1.18}
\definecolor{A}{HTML}{4DBD05}
\definecolor{B}{HTML}{1FBEB8}
\definecolor{C}{HTML}{0088C9}
\definecolor{D}{HTML}{6563C4}
\definecolor{E}{HTML}{9B45A3}

\usepackage{cite}
\usepackage{amsmath,amssymb,amsfonts}
\usepackage{algorithmic}
\usepackage{graphicx}
\usepackage{textcomp}
\usepackage{xcolor}
\def\BibTeX{{\rm B\kern-.05em{\sc i\kern-.025em b}\kern-.08em
    T\kern-.1667em\lower.7ex\hbox{E}\kern-.125emX}}

\makeatletter
\renewcommand\footnoterule{%
  \kern-3\p@
  \hrule\@width 1\columnwidth
  \kern2.6\p@}
  \makeatother
  
\begin{document}

\IEEEpubid{\makebox[\columnwidth]{ 
} \hspace{\columnsep}\makebox[\columnwidth]{ }}

\title{Critiquing Computing Artifacts through Programming Satirical Python Scripts}


\author{\IEEEauthorblockN{Aadarsh Padiyath}
\IEEEauthorblockA{\textit{School of Information} \\
\textit{University of Michigan}\\
Ann Arbor, USA \\
aadarsh@umich.edu}
\and
\IEEEauthorblockN{Tamara Nelson-Fromm}
\IEEEauthorblockA{\textit{School of Engineering} \\
\textit{University of Michigan}\\
Ann Arbor, USA \\
tamaranf@umich.edu}
\and
\IEEEauthorblockN{Barbara Ericson}
\IEEEauthorblockA{\textit{School of Information} \\
\textit{University of Michigan}\\
Ann Arbor, USA \\
barbarer@umich.edu}}

\maketitle

\begin{abstract}
Computing artifacts tend to exclude marginalized students, so we must create new methods to critique and change them. We studied the potential for ``satirical programming" to critique artifacts as part of culturally responsive computing (CRC) pedagogy. We conducted a one-hour session for three different BPC programs (N=51). We showed an example of a satirical Python script and taught elements of Python to create a script. Our findings suggest this method is a promising CRC pedagogical approach: 50\% of marginalized students worked together to create a satirical script, and 80\% enjoyed translating their ``glitches" into satirical Python scripts.\newline
\end{abstract}


\section{Introduction}

In its push to broaden participation in computing (BPC), the computer science (CS) community created several programs focused on increasing access and helping marginalized students succeed in a computing culture that tends to oust them \cite{gilbert2006making}. However, there are increasing calls for these programs to create and center methods for critiquing and changing the culture and its artifacts that left out marginalized students in the first place \cite{vakil2018ethics, ko2020time, guzdial2020cs}.

Scott et al. proposed the use of culturally responsive computing (CRC) and dialectical activities to broaden and sustain participation in computing, but more importantly to empower youth to see the flawed nature of the dominant culture \cite{scott2015culturally}. Additionally, activities with narrative and storytelling factors, such as satirical speculation \cite{klassen2022run}, have become increasingly popular and promising for both identity-centered and critical CS pedagogy \cite{jones2011interdisciplinary, shapiro2021using}. 

Many marginalized students have experienced what Dr. Ruha Benjamin calls ``glitches": a breakdown of computing artifacts regarding the intersection of identities and computing that expose systemic biases \cite{benjamin2019race}. She gives an example of the google maps narrator saying to turn onto ``Malcolm Ten Boulevard" rather than ``Malcolm X Boulevard." She says this is not a glitch but ``a form of evidence illuminating underlying flaws in a corrupted system." Storytelling through satire has the potential to draw attention to particular issues and use wit to expose and ridicule flaws in a system. Programming satirical scripts of ``glitches" have the unique possibility for marginalized programmers to learn and practice technical programming skills, expose systemic biases in software engineering, as well as reflect and demonstrate how technology intersects and interacts with their identities.

In this exploratory study, we describe the experience of using ``satirical programming" for scaffolding critical reflection of computing artifacts using culturally responsive computing methods. Specifically, we designed this study to answer the following research questions, based on tenets of CRC: 

\textbf{RQ1:} What are students self-reported perceptions when creating a satirical programming script about a ``glitch" they have experienced in a hour-long BPC session that extends beyond solely technical content knowledge?

\textbf{RQ2:} What percentage of these students use the ``glitch" approach in their satirical script?


We piloted this teaching method with majority students of color (N=51) in BPC programs: two from under-resourced high schools and one from community colleges in the Midwest region of the United States. This included first discussing and identifying how technology and computational artifacts interact with the student's identities, how programming elements in Python helped or hindered these interactions, and then creating a satirical narrative using the programming elements. To evaluate the research questions, we distributed a post-survey to measure reactions to the session and collected the students' Python programs for analysis.

\section{``Satirical Programming"}
\begin{figure}[h]
  \centering
  \includegraphics[width=\linewidth]{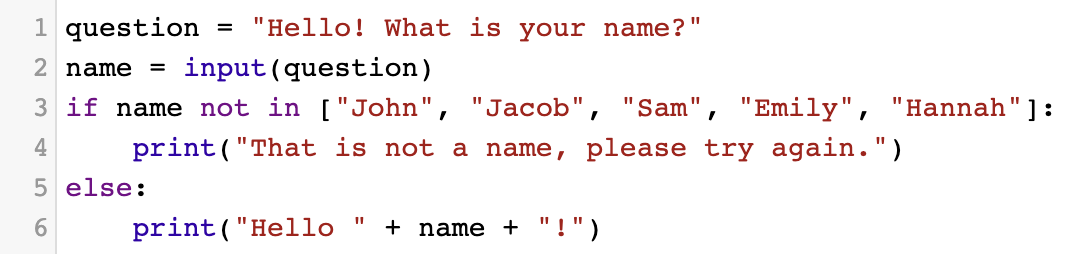}
  \caption{A satirical Python script of auto-correct based on the first author's experiences of typing a non-Western name in word processors.}
  \label{fig:autocorrect}
\end{figure}

In this study, we define the term \textit{``Satirical Programming"} as a programming script that uses satire/comedy to express a social critique, often of systemic biases. For example, when writing the first author's traditionally non-Western name on a word processing document, it frequently will be auto-corrected, but a traditionally Western name will not. In this situation, we can see how auto-correction technology was not designed for names like the first author's. This bias against non-Western names can be shown through a satirical programming script as shown in Figure \ref{fig:autocorrect}.

\section{Literature Review}

Recognizing its part in global social injustice and systemic racism as well as its non-representational population of students with marginalized identities, the CS community created several programs for broadening participation in computing. These programs can be crucial for increasing access and helping marginalized students endure computing culture. However, there are increasing calls for these programs to center critiquing and changing the culture and its artifacts that ousted marginalized students in the first place \cite{ko2020time, guzdial2020cs}. Additionally, this refocusing towards systemic issues in computer science can help sustain participation in computing, as this aligns with the pro-social goals of students who identify as women, first-generation college students, Black/African-American, Hispanic/Latinx/Latin* and/or LGBTQIA+ \cite{lewis2019alignment, denner2015computer, stout2016lesbian}. This is especially true for students with intersectional identities such as Black women, who often don't wish to reinforce these oppressive systems \cite{rankin2020intersectional}. Thus, we must begin to explore new pedagogical methods for computer science that center critiquing and changing the culture and artifacts of CS.

\subsection{Culturally Responsive Computing Education}

In a 2015 review of the over 50 US-based BPC programs for raced-gendered-ethnic minority students, Scott and Clark found the majority focused solely on issues of technical literacy and did not mention issues of diversity, culture, or identity \cite{scott2015culturally}. This stands in direct contrast to education research literature regarding culturally responsive and sustaining teaching methods for increasing academic achievement, self-efficacy, and belonging among minoritized youth \cite{gay2013culturally, paris2012culturally, howard2011culturally,rorrer2018national, codding2019positionality}. Culturally responsive pedagogy uses students' diverse identities, cultures, and experiences as the basis and vehicle for learning.

Scott et al. reconceptualized this theory in the context of computing education \cite{scott2015culturally}. The authors proposed five tenets of culturally responsive computing (CRC): (1) All students are capable of digital innovation; (2) The learning context supports transformational use of technology; (3) Learning about one’s self along various intersecting sociocultural lines allows for technical innovation; (4) Technology should be a vehicle by which students reflect and demonstrate understanding of their intersectional identities; and (5) Barometers for technological success should consider who creates, for whom, and to what ends rather than who endures socially and culturally irrelevant curriculum. We centered these tenets in the design of our learning session. Examples of CRC include contextualizing a computing curriculum around topics of social justice \cite{niaseducational} and relating computing concepts to Historically Black Colleges and Universities (HBCU) culture \cite{washington2021opportunities}. 

\subsection{Creative Methods for Critiques in Computing Education}

In our shift to culturally responsive pedagogy in computing, we must also explore new pedagogical methods that center critiquing and changing the culture and artifacts of CS. The computing education community has had several recent innovations in using creative methods for eliciting critiques of computing. Scholars in ethical and critical computing specifically use creative storytelling factors such as visual metaphors to access and express students' understanding of structural problems in computing \cite{kirdani2022house} and satirical ethical speculation as a probe to excite students and discuss the role of computing in society \cite{klassen2022run}.

Kirdani-Ryan and Ko introduced critiques of computing alongside technical computing content and assessed students' understanding through the artistic creation of ``floorplans": visual metaphors for representing systems and problems in the field of computing \cite{kirdani2022house}. In designing the floorplans, every student was successful in connecting their technical content with structural issues in computing and its applications. For example, on one student's floorplan, they drew an image representing a large ``the old women's bathroom" which was torn down in the 1970s to make room for a men's bathroom; perhaps remarking on the shift in computing away from being a ``women's career." This work provided evidence that students can successfully generate computing critiques using artistic and creative methods. The authors provide several takeaways from the research, including a suggestion of a framework or scaffolding for approaching creative work as many students had difficulties in learning the new assignment structure. We build on this work through programming creative critiques in an introductory computing lesson and take the authors' suggestion of providing scaffolding for students' Python scripts.

Klassen and Fiesler introduced a tool for scaffolding critical and ethical speculation through a classroom activity called the ``Black Mirror Writers Room" \cite{klassen2022run}. The activity is inspired by \textit{Black Mirror}, a British anthology television series that uses satire and speculative fiction by showing exaggerated near-future dystopias inspired by our current relationships with technology. The activity involved students working in small groups to choose an issue or technology; answer questions speculating futures of the technology, specifically its possible harms or negative outcomes; discuss whose story could best present cautions that technologists should consider; and create an episode blurb that pitches the concept to the \textit{Black Mirror} writers \cite{klassen2022run}. The activity was extremely well received by the students and instructors and gave students, even students in non-technical classes, the opportunity to engage with issues of ethics and technology through a fun and engaging lens. One challenge the authors suggest addressing is to scaffold discussions of ethics and critiques in non-ethics classes. We build on this work through the use of generating satirical critiques of computing artifacts with programming and use the authors' suggestion of scaffolding the critiquing process.

This use of satire may draw doubt. Responding comedically to oppressive technologies may imply an indifference or un-seriousness towards what are deeply serious problems. However, ethnographic research with marginalized youth, specifically queer youth of color, find the populations using humor in response to oppression in order to practice critique, elicit pleasure, and subvert hatred \cite{blackburn2005agency, shrodes2021humor}. The use of satire as a method for critique establishes the youth as agents that resist oppression and sustain joy through the creation of comedic critiques that are still grounded in the seriousness of the problem \cite{shrodes2021humor}. We extend this scholarship by exploring the possibility of satire as a method for critique in programming.

The works of Kirdani-Ryan and Ko \cite{kirdani2022house} as well as Klassen and Fiesler \cite{klassen2022run} served as inspiration for us as we investigated the intersection of the works using culturally responsive pedagogy: programming satirical scripts to draw attention to ``glitches" marginalized students face that are emblematic of systemic issues in software engineering.

Outside of ethics and critical computing, introducing programming through creativity has been promising in attracting and retaining diverse students  \cite{groeneveld2022creatively}. Additionally, a major theme in this literature, as stated by Jones et al., is to put "creativity first, programming second" \cite{jones2011interdisciplinary}. We reiterated this theme throughout our session, urging students to focus on their creativity and story, rather than on programming.

\subsection{Designing the Session}

There were many considerations in designing a learning session to scaffold students in writing satirical Python scripts while learning technical coding skills. Previous literature contributes several lessons, concerns, and guidelines when guiding culturally responsive computing education.


As Klassen and Fiesler suggest, scaffolding the critiquing process is beneficial for discussions of ethics and critiques \cite{klassen2022run}. Freire suggests discussing students' \textit{limiting situation}: a (often historical) condition that restricts one's freedom, for another group's benefit \cite{freire2018pedagogy}. We began the probe by discussing some generic limiting situations with software that students may be interested in such as automated software for admitting students into college via GPA. We use this limiting situation to explain hidden assumptions in this system. For example, if a student lost a loved one in a semester, and this negatively influenced their GPA, a simplistic algorithm would penalize the student for a situation out of their control. The instructor also shared their limiting situations with software, as Kirdani-Ryan et al. wrote in their experience that personal situations were the most expressive and powerful \cite{kirdani2022house}. For example, when writing my traditionally non-Western name on a word processing document, it frequently will be auto-corrected, but a traditionally Western name will not. In this limiting situation, we can see how auto-correction technology was not designed for names like mine. After sharing this information, we made space for students to reflect and share similar limiting situations of how technology interacts with and perhaps doesn't account for their own identities. This process can be scaffolded by asking students to think of times when software couldn't understand them. After writing down and sharing these experiences, we can connect them to Dr. Ruha Benjamin's concept of ``glitches:" a breakdown of computing artifacts regarding the intersection of identities and computing that expose systemic biases. By engaging with generic limiting situations and the lecturer's situations, we can prepare students to notice and explain the prevalence of these situations.

After discussing limiting situations, we can shift this critical lens towards the programming language Python. By connecting the assumptions and rules of interacting with technology to syntax in Python, we can approach Python with a lens of skepticism as well as focus on the hidden rules and assumptions the creator of Python, Guido van Rossum, imbued into the language \cite{format_2005}. This can highlight who and what Python was designed for, the implicit assumptions baked into the language, and positions students as critical observers and users of the language \cite{ko2022critically}.

We then shift back to our limiting situations, as we can now apply the newfound knowledge of Python to create satirical scripts. However, creative expression may be difficult for STEM students. Kirdani-Ryan et al. found that students struggled with creative expression and requested a framework or further scaffolding \cite{kirdani2022house}. We can use the personal limiting situations with software discussed previously to scaffold the counternarratives. They can choose a limiting situation that they relate to or one that speaks to them and tell an interactive story through the elements of Python they learned in the session. We also provided an example to clarify the prompt. The example communicated the first author's relationship with auto-correct in a satirical Python script, shown in Figure \ref{fig:autocorrect}. Additionally, students can work in groups to write the Python scripts together. This option will help reluctant students support each other with the probe and support each other's expression through computing.

\section{Description of Practice}

\subsection{Learning Session}

We created a one-hour session outline as a probe for a lecture session in three BPC programs in the American Midwest based on findings from a review of the literature. This probe is inspired by and uses concepts from ``Chapter 4: Teaching CS, Equity, and Justice" and ``Chapter 10: Programming Languages" in \textit{Critically Conscious Computing} \cite{ko2022critically}.

\noindent \textit{Learning Objectives:}

\begin{enumerate}
    \item Students will be able to use variables, print strings, and use the input function in the Python programming language.
    
    \item Students will be able to create a satirical Python script about a ``glitch" they experienced.
    
\end{enumerate}

\noindent \textit{Session Outline:}

\begin{itemize}
    \item (5 mins) Begin the session by discussing rules in social systems, how assumptions are inherent to these rules, and how it translates to the design of software. Support the concept by starting the discussion with assumptions inherent to a context they may be familiar with: college admissions.
    \item (10 mins) Ask students to help brainstorm in groups the many situations in which their identity was not accounted for by software. Scaffold this through thinking about times technology didn't understand them, or through popular things they heard about in the news. As students brainstorm, write these ``glitches" down in a shared space so that everyone can see them.
    \item (5 mins) Explain that software is designed by a specific person or group of people with particular experiences and values. The negative experiences that students have with software are a result of a mismatch of these experiences and values. Additionally, as the group of people that design software is overwhelmingly White and Asian men, the software often ``glitches" when encountering other identities, exposing systemic biases.
    \item (5 mins) Discuss the relationship between rules and computing languages: This includes syntax of strings, variables, and the input and print function in Python as decisions the designer of Python made.
    \item (15 mins) Walk through practice problems for the Python concepts, with a focus on the hidden assumptions of using Python. For example, this can include a discussion of the knowledge barrier of error messages in Python. When omitting a closing quote on a string, the compiler throws the error: ``Syntax Error: EOL while scanning string literal." This information is not accessible to new programmers and contains programmer jargon. This can be connected to the assumptions made by the designer of Python: programmers using Python will be able to understand this message and have enough information to fix the issue.
    \item (15 mins) Ask students to choose a ``glitch" they or their classmates shared, specifically one they would like to discuss or one that impacts themselves. Showcase a personal experience of a ``glitch" turned into a satirical script. Then discuss together how they can tell an interactive story about the experience by using strings, variables, and the input function in Python. They can work together in groups if they like, or they can work by themselves.
    \item (5 mins) Allow students the opportunity to share their creations for everyone to view if they wish to share. Then end the session by bringing assumptions and software back into focus. Discuss the importance of listening to those that are often marginalized, and working with marginalized communities to design software for a world we all belong in.
\end{itemize}

\subsection{Sites}

We conducted this study with three BPC Summer programs for high school or community college students entering college computing programs: 
Wolverine Pathways (WP)\footnote{https://wolverinepathways.umich.edu/}, LSAMP NxtGEN STEM Scholars Program (LSAMP)\footnote{https://www.milsamp.org/programs/}, and Community College Summer Institute (CCSI)\footnote{https://www.si.umich.edu/programs/bachelor-science-information/new-transfers/umsi-community-college-summer-institute}
. All three programs have goals of exposing underrepresented ethnicity and socioeconomic-status students to undergraduate level lectures and courses at 
the School of Information at the University of Michigan, Ann Arbor
. This includes seventh through twelfth-grade students from under-resourced high schools in large school districts in the state of Michigan (
WP
); STEM-interested high school students from backgrounds underrepresented in computing (URM): African Americans, Hispanic Americans, American Indians, Alaska Natives, Native Hawaiians, and Native Pacific Islanders (
LSAMP
); and Michigan community college students (
CCSI
). Students were recruited through emailing lists to STEM teachers from under-resourced high schools (
WP, LSAMP
), as well as to emailing lists to community college students (
CCSI
). The 
CCSI
Session was conducted in person, and 
WP and LSAMP
were conducted over Zoom video-conferencing software. A breakdown of reported student demographics is displayed in Table \ref{tab:demo}. Each site had one instructor (the first author) use the session outline described in section 3. Two sessions were 1 hour long (
CCSI, LSAMP
), and one session was 2 hours long with substantial breaks throughout for 1 hour of total content (
WP
). Before each session, students spent an hour on ice-breakers and assorted group-work. After the 
LSAMP and WP
sessions, the students were dismissed for the day. After the
CCSI 
session, the students had one more hour-long session before they were dismissed for the day.

\begin{table}[ht]
    \centering
    \begin{tabular}{|c||c|c|c|}
        \hline
        \textbf{Program} & WP & LSAMP & CCSI \\
        \hline
        \textbf{\# of Participants} & 15 & 13 & 23 \\
        \hline
        \textbf{Gender Ratios} & \makecell{60\% Women\\40\% Men} & \makecell{38\% Women\\62\% Men} & \makecell{57\% Women\\43\% Men} \\
        \hline
        \textbf{URM Ratio} & 95\% & 100\% & 87\% \\
        \hline
        \textbf{Delivery} & Zoom & Zoom & In-person \\
        \hline
    \end{tabular}
    \newline
    \caption{Demographic breakdowns of each site retrieved by 
    University of Michigan's
    DEI Office.}
    \label{tab:demo}
\end{table}

\subsection{Data Collection}

Parental consent and student assent was collected before each session as requested by the 
University of Michigan
IRB. However, if a student was over the age of 18, they provided their own consent. Students were informed that their learning experience as well as any artifacts created may be collected.

\subsubsection{Post-survey}

To understand students' reactions to this learning session, we asked the following questions in an online survey after completing the session. The survey was sent to the email address provided to the program right after the session. These were often either the student's or their parent's email address. At the end of each session the instructor asked students to look in their email inbox and fill out the survey. This survey uses a 5-point Likert scale for each statement ranging from 1 = ``I hated it", 2 = ``I disliked it", 3 = ``It was okay", 4 - ``I liked it", to 5 = ``I loved it". This survey context was modified to fit this study from an NCWIT E-Textiles Teacher Workshop Experience Survey that evaluated teachers' experience with an introductory computing workshop, including enjoyment \cite{ncwitetextileteacherworkshopurvey}.

\begin{enumerate}
    \item Session Overall
    \item Brainstorming ``Glitches" Part of Session
    \item Learning Programming Part of Session
    \item Writing Stories Part of Session
\end{enumerate}

Additionally, we measured the difficulty and pace of the session. We asked the following questions with a 3-point Likert scale for each statement along with an open-response area for further elaboration.

\begin{enumerate}
    \item Level of difficulty, ranging from 1 = ``Too hard" to 2 = ``Just right", to 3 = ``Too easy"
    \item Pace of the workshop, ranging from 1 = ``Too Fast" to 2 = ``Just right", to 3 = ``Too slow"
\end{enumerate}

Lastly, we had a final open-ended question: \textit{Please tell us about your experience during the session. What was fun, what was challenging, what was frustrating?}

\subsubsection{Runestone Logging}

The course sessions used the interactive open-source e-book system, Runestone, as a programming environment \cite{ericson2020runestone}. Students were presented with a Runestone assignment page containing practice problems relating to the technical concepts in class, as well as space to write their satirical script. Coding logs and satirical Python scripts were collected through this system. Specifically, it collected submissions to the Python compiler when attempting to run written code.

\subsection{Qualitative Analysis}
To compile themes from students' ``glitches," Python scripts, and the post-survey open-ended responses, we used the thematic analysis coding approach provided by Braun et al. \cite{braun2006using}. This involved coders generating codes from the data independently. These codes were then grouped into themes. The themes were discussed and agreed upon by the two coders. All ``glitches" and satirical Python scripts were then fully thematically coded by both coders simultaneously such that both coders agreed on each code. Thus we forgo reporting inter-rater reliability as guided by the arguments of Hammer and Berland \cite{hammer2014confusing}. There were minimal disagreements in the thematic coding process, with disagreements primarily regarding the specificity of themes. We mutually decided to employ more precise thematic codes to reduce mis-interpretations.

\section{Positionality}

Key to culturally responsive computing pedagogy is the need for cultural competence of educators \cite{washington2020twice}. As this requires serious and continuous reflection of an educator's identities and motivations, I share how my experiences led to and influenced this study. Several experiences of my identity clashing with computing culture and its artifacts led me to rejecting the harmful dominant ideologies of computing culture and heavily informed this work. I do this work to critique and change computing culture and its artifacts to create futures we can live and belong in. As a \textit{Desi} (South Asian) man, my identity stands within a prominent subgroup of the dominant culture of computing: cis-gendered White and Asian men. Although my queer identity is certainly not the norm within this dominant culture, there is still a challenge and gap in research in bringing my identity and positionality to this context. Many White lecturers within the dominant culture of education struggle to engage in discussions of critical counternarratives, race, and power \cite{johnson2011road, porfilio2011guiding, bolgatz2005teachers}. Thus I approach this context with humility and care.

\section{Experiences and Reflections}

\subsection{Student Artifacts and Reactions}

In analyzing students' (N=51) experiences, we found that several students exhibited existing or newfound recognition of ``glitches" that revealed systemic biases encoded in technologies. Students worked in groups to discuss and share ``glitches" they have experienced, submitting a total of 23 ``glitches." Repeating themes in students' written experiences included non-traditional names and auto-correct (8/23; 35\%); light-skin bias in cameras/camera filters (4/23; 17\%); White hairstyle bias in search results and video games (3/23; 13\%); and restrictive or automatic categorizations (4/23; 17\%). Examples of student responses grouped by these themes are displayed in Table \ref{tab:glitches}.

\begin{table}[h]
    \centering
    \begin{tabular}{|p{0.3\linewidth} | p{0.6\linewidth}|}
    \hline
        Non-traditional names and auto-correct & When I search my name in a
search engine, the meaning always
comes up as ``tree" when my
parents always tells me it means
god's message. \\
\cline{2-2}
         & When I type my name it can
auto-correct to a different name,
and many people mishear me
when I say it to them, which leads
to them recording it wrong. \\
        \hline
        Light-skin bias in cameras/camera filters & My friend who is of darker skin
color than mine, was having
issues getting software to
recognize her face and pattern it
correctly \\
\cline{2-2}
         & Often times, photo editing
software or phones will tell darker
skinned people to turn up
exposure or won't recognize their
face at all\\
        \hline
        White hairstyle bias in search results and video games & In most video games, most
of the preset game hairstyles are
straight or only have Euro-centric
hairstyles.\\
\cline{2-2}
         & When looking up ideas, such as
makeup, hairstyles or outfit ideas
the search engine assumes our
race and we never see people of
color.\\
        \hline
        Restrictive or automatic categorizations & Automatic gender
assumption cookie when online shopping it
automatically picks a gender
specific clothing. \\
\cline{2-2}
         & Race options on
applications: Being mixed raced and not
knowing what option to choose, or
the reason for needing to specify
race on an application at all\\
        \hline
    \end{tabular}
    \newline
    \caption{Example student responses of ``glitches" in their interactions with technology grouped by repeating themes.}
    \label{tab:glitches}
\end{table}

\begin{figure}[h]
  \centering
  \includegraphics[width=\linewidth]{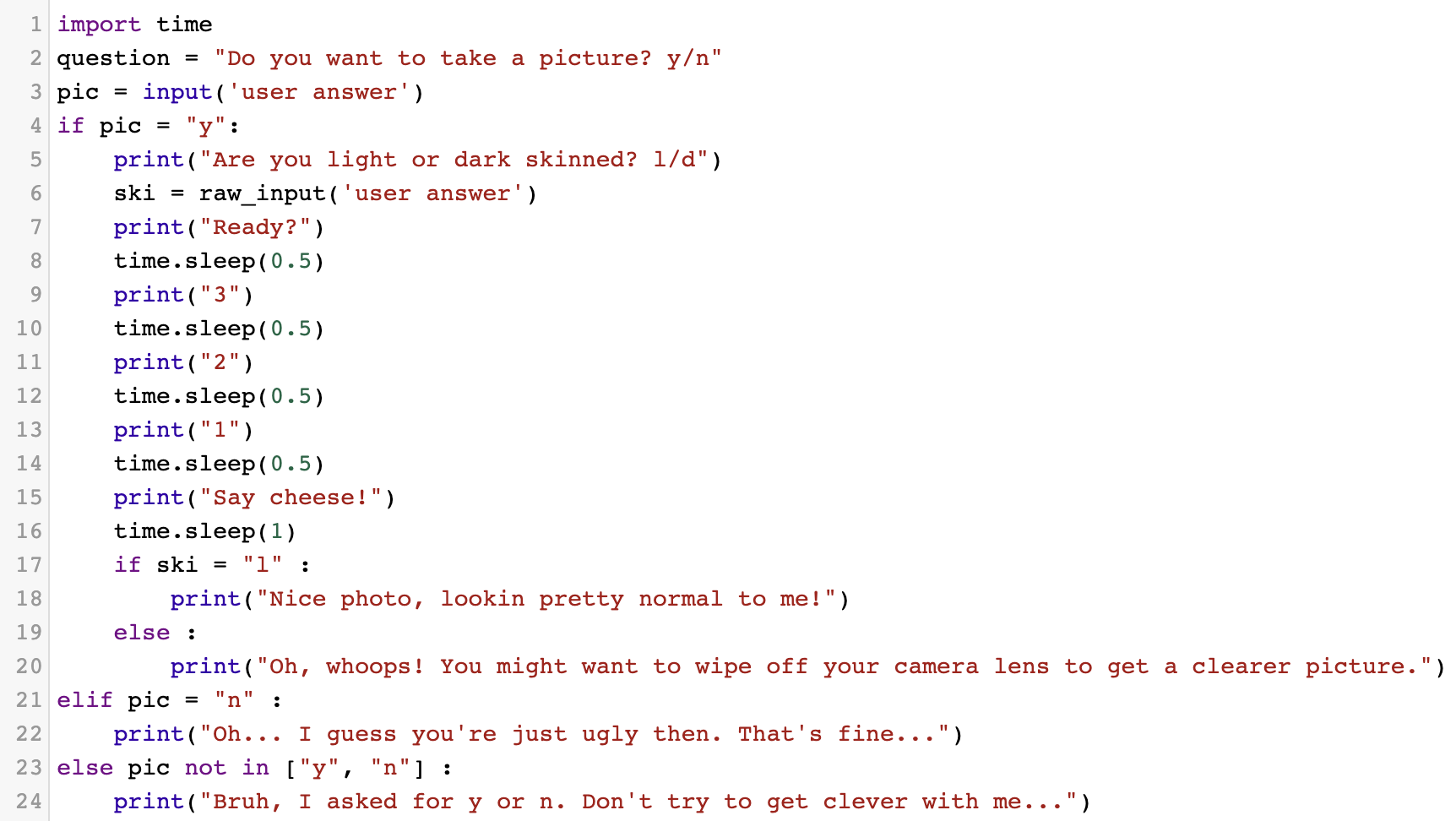}
  \caption{A satirical Python script of using a camera based on a student's experience with camera filters.}
  \label{fig:camera}
\end{figure}

Many students were able to translate these ``glitches" into critiques in the form of satirical Python scripts. 
One student created a satirical critique of their experience with a light-skinned bias in cameras and camera filters. The script, shown in Figure \ref{fig:camera}, asks if a user wishes to take a photo. If the user does, it then asks if the user is light or dark skinned. If they are light-skinned, the script prints ``Nice photo! Lookin pretty normal to me!" However, if they are dark-skinned, the script prints, ``Oh whoops! You might want to wipe off your camera lens to get a clearer picture."

Another student translated their experience with White bias in recommendations for hairstyles into a satirical critique. The script first notes that it will provide a hairstyle recommendation based on someone's name. The script then asks the user for a name. It compares the name against 5 common Western White women names. If the name was one of those five names, it recommended ``Hair in messy bun." Otherwise, for all other names, it recommends ``Cornrows."

\begin{figure}[h]
  \centering
  \includegraphics[width=\linewidth]{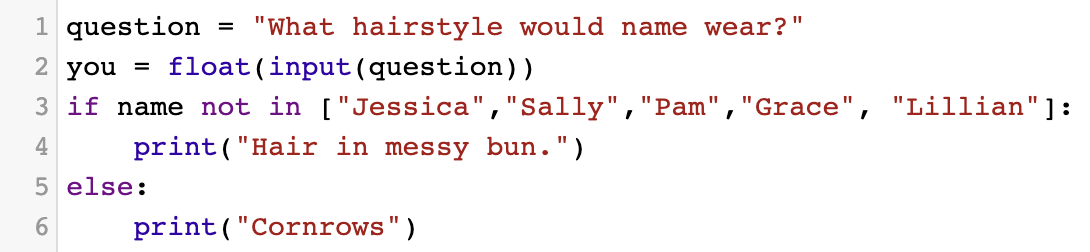}
  \caption{A satirical Python script of hairstyle recommendations based on a student's experience with online search results.}
  \label{fig:hairstyle}
\end{figure}

\begin{figure}[h]
  \centering
  \includegraphics[width=\linewidth]{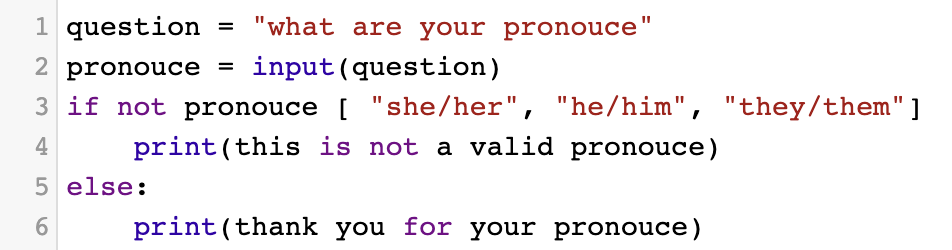}
  \caption{A Python script to validate pronoun options.}
  \label{fig:pronouns}
\end{figure}

Students also worked in groups or by themselves to submit a total of 34 Python scripts. In 7 out of 34 scripts (21\%), students were able to identify a ``glitch" in their experiences with technology and communicate a satirical critique of the experience in a Python script. Many students (10/34; 29\%) created the beginnings of a critique through their script, however, the objective of their script was unclear or missing. For example, the intention of the script in Figure \ref{fig:pronouns} is unclear. The student could be making a statement about many gender and pronoun related issues regarding their intersections with technology, such as a lack of fluidity or intersections in options (such as they/she), a lack of neopronoun options (such as xe/xem), and more. However, this ambiguity reduces the ability for the student to communicate a specific and clear critique with this script.

\begin{figure}[h]
  \centering
  \includegraphics[width=\linewidth]{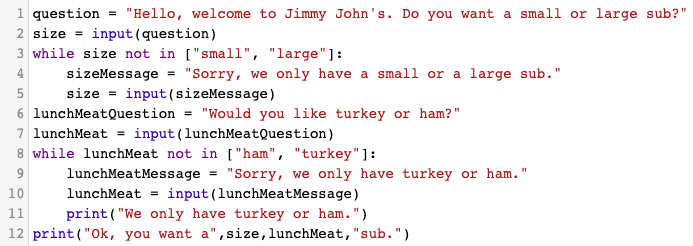}
  \caption{A Python script to order a sandwich from Jimmy John's.}
  \label{fig:jimmy}
\end{figure}

In 17 out of the 34 Python scripts, the students chose to not create a satirical critique, but rather express themselves through creating an interactive script. For example, in Figure \ref{fig:jimmy} a student created a script for ordering a sandwich from Jimmy John's.

\begin{figure}[h]
    \centering
    \begin{tikzpicture}
  \begin{axis}[
    height = 3cm,
    width = 6cm,
    xbar,
    x axis line style = { opacity = 0 },
    hide x axis,
    tickwidth         = 0pt,
    bar width = 5pt,
    enlarge y limits  = 0.15,
    enlarge x limits  = 0.08,
    symbolic y coords = {Level of Difficulty},
    ytick=data,
    nodes near coords,
    legend style={at={(1.0,1.03)},anchor=south east},legend columns=5,
    table/col sep=comma
  ]
  \addplot[color=A, fill=A] 
	coordinates {(3,Level of Difficulty)};
\addplot[color=C, fill=C]  
	coordinates {(21,Level of Difficulty)};
\addplot[color=E, fill=E]  
	coordinates {(1,Level of Difficulty)};
\legend{Too Hard,Just Right,Too Easy}
  \end{axis}
\end{tikzpicture}
\begin{tikzpicture}
  \begin{axis}[
    height = 3cm,
    width = 6cm,
    xbar,
    x axis line style = { opacity = 0 },
    hide x axis,
    tickwidth         = 0pt,
    bar width = 5pt,
    enlarge y limits  = 0.15,
    enlarge x limits  = 0.08,
    symbolic y coords = {Pace of the workshop},
    ytick=data,
    nodes near coords,
    legend style={at={(1.0,1.03)},anchor=south east},legend columns=5,
    table/col sep=comma
  ]
  \addplot[color=A, fill=A] 
	coordinates {(2,Pace of the workshop)};
\addplot[color=C, fill=C]  
	coordinates {(23,Pace of the workshop)};
\addplot[color=E, fill=E]  
	coordinates {(0,Pace of the workshop)};
\legend{Too Fast,Just Right,Too Slow}
  \end{axis}
\end{tikzpicture}
    \caption{Distribution of responses (N=25) to the post-survey containing a 3-point Likert scale question regarding perceptions of the session difficulty and pace.}
    \label{fig:difficulty_pace}
\end{figure}
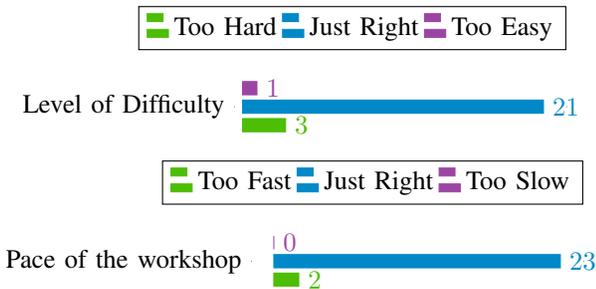

Students' reactions to the session were very positive. Based on the post-survey (n=25/51, 49\% response rate), 80\% of students responded positively (4/5 - ``I liked it," or 5/5 - ``I loved it") to the session overall. Figure \ref{fig:likert_questions} displays the distribution of the reactions to the session and its parts. A majority of students ``liked" or ``loved" every part of the session, including the session overall. 84\% of students felt the session was appropriately paced (21/25 rated ``Just Right") and 92\% of students appropriately difficult (23/25 rated ``Just Right"). Figure \ref{fig:difficulty_pace} display the distribution of the perceptions of the session difficulty and pace.

\begin{figure}
\begin{tikzpicture}
\pgfplotsset{%
    width=.7\linewidth,
    height=7.5cm
}
\begin{axis}[
    xbar,
    x axis line style = { opacity = 0 },
    hide x axis,
    tickwidth         = 0pt,
    bar width = 6pt,
    enlarge y limits  = 0.15,
    enlarge x limits  = 0.08,
    symbolic y coords = {Writing Stories,Learning Programming,Brainstorming ``Glitches",Session Overall},
    ytick=data,
    nodes near coords,
    legend style={at={(0.8,1.03)},anchor=south east},
    legend columns=3,
    table/col sep=comma
]
\addplot[color=A, fill=A] 
	coordinates {(0,Session Overall) (0,Brainstorming ``Glitches")
		 (0,Learning Programming) (0,Writing Stories)};
\addplot[color=B, fill=B]  
	coordinates {(1,Session Overall) (3,Brainstorming ``Glitches")
		 (0,Learning Programming) (1,Writing Stories)};
\addplot[color=C, fill=C]  
	coordinates {(4,Session Overall) (5,Brainstorming ``Glitches")
		 (7,Learning Programming) (8,Writing Stories)};
\addplot[color=D, fill=D]  
	coordinates {(14,Session Overall) (14,Brainstorming ``Glitches")
		 (12,Learning Programming) (12,Writing Stories)};
\addplot[color=E, fill=E]  
	coordinates {(6,Session Overall) (3,Brainstorming ``Glitches")
		 (6,Learning Programming) (4,Writing Stories)};
\legend{I hated it,I disliked it,It was ok,I liked it,I loved it}
\end{axis}
\end{tikzpicture}
\caption{Distribution of responses (N=25) to the post-survey containing 5-point Likert scale questions regarding reactions to the session and its parts.}
\label{fig:likert_questions}
\end{figure}
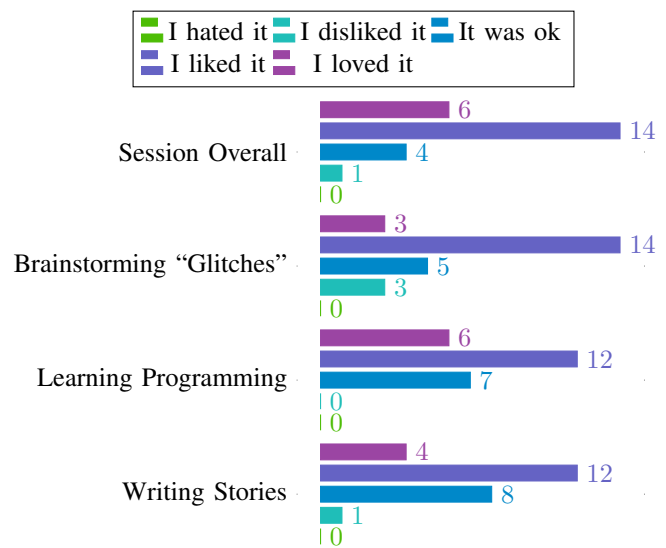

In response to the open-ended question in the post-survey, a majority of students (n=10/18; 56\%) specifically appreciated the critical approach to Python. One student compared this session to a previous experience attempting to learn Python:

\begin{quote}
    \textit{I enjoyed seeing a different approach to Python because I've used it before and had a less than ideal experience.}
\end{quote}

Another mentioned a newfound awareness of how systemic biases are programmed into software:

\begin{quote}
    \textit{I liked how it made me think about programming in a different way and how there could be internal bias from the programmer.}
\end{quote}

Several students noted having fun translating their experiences into a Python script, noting that this challenge kept the learning experience fun (12/18; 67\%):

\begin{quote}
   \textit{I found it fun because I personally enjoy to code and I found it fun to create scenarios with code as it was creative and teaching at the same time.}
\end{quote}

Several students mentioned the experience was challenging yet fun and engaging (7/18; 39\%). One student responded:

\begin{quote}
    \textit{Brainstorming was fun, but figuring out how to word the story in the code was challenging for me.}
\end{quote}

Another student had difficulty in the brainstorming section of the session:

\begin{quote}
    \textit{I found getting to create my own code really fun. I was having issues coming up with ideas to write for my story. I didn't really find anything frustrating.}
\end{quote}

One student response spoke of confusion in some points of the session, however they found working with others enjoyable and helpful:

\begin{quote}
    \textit{At some parts I got confused, but as you go on and work with others, it gets much easier. It's very fun working with people.}
\end{quote}

Some students disliked parts of the session and one student disliked the session overall according to the post-survey. Two students that disliked the ``Brainstorming Glitches" section wrote:

\begin{quote}
    \textit{Sometimes I was trying to make something and it was taking me a little while and then before I was done we were moved on.}
\end{quote}

\begin{quote}
    \textit{Overall it was fun and very engaging, I just dont like zoom :(}
\end{quote}

The other students that rated ``I disliked it" to sections opted not to respond to the open-ended question.

\subsection{Instructor Reflections}
Throughout the design and execution process, I actively worked to hold myself accountable for any possible harm caused by this research. This included a thorough self-reflection of my values and motivation regarding this project and my positionality; seeking out minoritized individuals from similar backgrounds of the participants whom I trusted to give me honest criticism of this study; explicitly asking for pushback if students felt uncomfortable or were apprehensive throughout the session; vigilantly observing in-person student reactions for uneasiness or wariness throughout the sessions; and planning for divergences from the session plans if necessary. I recommend this preparation and vigilance for those looking to adopt this method, as this approach is not suitable for all contexts.

In the design of this session, we expected difficulties in both identifying ``glitches" and translating them into satirical programs. Most students recognized ``glitches" in their experiences with technology, however, many students still had difficulties in translating their experiences into a focused critique within a Python script. This was challenging, as there were several groups of students working on Python scripts, but only one instructor to assist them. When students had difficulties in proposing a satirical critique and asked for assistance, we focused on identifying the systemic biases the ``glitch" exposes. Then, we discussed what experiencing the ``glitch" felt like, and what seemed absurd or stupid about it. Next, we urged the students to re-create that absurd experience using Python to communicate their satirical critique.

Although we stressed ``creativity first, programming second" throughout the session, several students were frustrated when their code had errors. In assisting these students, we reminded the students of the unhelpful error messages the creator of Python designed. This shifted the blame of not understanding Python from the students to the designers of Python.

\section{Discussion}
In this study, we investigated the use of ``satirical programming" in scaffolding critical reflection of computing artifacts. We evaluated a one-hour learning session through a post-survey and an analysis of the satirical Python scripts. We discuss the results to our research questions in the following sections.

\textit{RQ1: What are students self-reported perceptions when creating a satirical programming script about a ``glitch" they have experienced in a hour-long BPC session that extends beyond solely technical content knowledge?}

A key tenet of CRC, as defined by Scott et al., is to ensure the learning context supports transformational use of technologies; in other words, the learning context should extend beyond transferring solely technical content knowledge \cite{scott2015culturally}. We centered this tenet through teaching a critical angle of Python and facilitating discussion on how biases are imbued into the design of software. The success of this was shown in the majority positive Likert-scale responses to every part of the session (Brainstorming ``Glitches": 68\% positive; Learning Programming 72\% positive; Writing Stories: 64\% positive), including the session overall (80\% positive) as shown in Figure \ref{fig:likert_questions}. In the open-ended question in the post-survey, a majority (56\%) of students specifically mentioned enjoying the critical approach to Python. Additionally, most (67\%) students noted having fun programming the satirical Python scripts. This all shows great promise for this method in supporting and exciting marginalized students through this transformational use of technologies. However, similar to previous research in this area \cite{klassen2022run, kirdani2022house}, students found creative expression in computing contexts challenging and required substantial scaffolding. Students may have disliked sections or the session overall due to the pace, or to the Zoom context.

\textit{RQ2: What percentage of these students use the ``glitch" approach in their satirical script?}

Another tenet of CRC is to position technology as a vehicle by which students reflect and demonstrate understanding of their intersectional identities \cite{scott2015culturally}. We centered this tenet through explicitly discussing students' experiences with technology, and translating their ``glitches" into satirical Python scripts. We found the use of the ``glitch" metaphor created by Dr. Ruha Benjamin \cite{benjamin2019race} to be very helpful in scaffolding the identification of systemic biases. 

Students readily accepted and applied the ``glitch" metaphor to identify limiting situations in their experiences with technology. We can see this through their submitted ``glitches" and subsequently the satirical Python scripts. The percentage of students that used the ``glitch" approach in their satirical script was 50\% (17/34). This method showed promise in affording these students' reflection and understanding of how their experiences with technology are influenced by their identities, through understanding how systemic biases manifest in software. When helping students write scripts, several students were unsure how to turn their ``glitches" into interactive satirical Python scripts. Students who had difficulty with this translation and did not ask for help may have felt more comfortable creating an interactive story rather than a satirical script. This suggests the need for further support in how to turn their ``glitch" into a satirical script. 


\subsection{Limitations}

There are several key limitations to this study. One limitation is the context in which the project was given: this study was conducted with a relatively small number of participants, with a specific demographic of participants, at one school, in one institution and in one country. The demographics of the students were unable to be connected to specific responses due to a IRB restriction, preventing intersectional analysis. The response rate for the post-survey was low, possibly due to the programs sending the survey to parents' email addresses. The survey used is not validated, so the results may be subject to measurement error. Only 18 out of 51 students (35\%) responded to the open-ended question in the post-survey. Students may have realized I was a researcher and responded positively to the survey to please me. Additionally, this was the first author's first teaching experience. Instructors with more teaching experience may find this method easier to employ.

\subsection{Future Work}

Future work can address these limitations and explore the pedagogical methods and metaphors used in this study. For example, more research is needed in employing this method in classes with diverse demographics. The use of satire in programming with diverse classes could support cross-cultural understanding, or alternatively could place undue burden and hypervisibility on minority students.

\section{Conclusion}

In this study we explored a pedagogical method for identifying systemic biases in computing artifacts and creating satirical critiques of our experiences. Our findings show that this method was promising for culturally responsive computing classes in that (RQ1) students enjoyed the session beyond solely the technical content knowledge and (RQ2) enabled several students to understand "glitches" and create satirical scripts that reflects how their identities interact with technologies. As students lives increasingly intersect with technology, minoritized youth deserve to understand how systemic biases manifest in software and how to push forward from oppressive experiences. Satirical programming scripts are a unique rhetorical device positioned to turn these oppressive experiences with technology into constructive and powerful critiques about the necessity for diversity, equity, and inclusion in software development.

\section*{Acknowledgments}
Thanks to Mark Guzdial and Patricia Garcia for valuable feedback. Thanks to Tamara Nelson-Fromm for assisting in qualitative coding.

\bibliographystyle{IEEEtran}
\bibliography{sources.bib}

\end{document}